\documentclass[useAMS,usenatbib]{mn2e}

\usepackage{txfonts}
\usepackage[dvips]{graphicx}
\usepackage{epsfig}
\usepackage{natbib}

\usepackage{multirow}
\usepackage{url}
\bibpunct{(}{)}{;}{a}{}{,}
\usepackage{color}
\title[Angular momentum evolution in haloes]%
{Angular momentum evolution in Dark Matter haloes: a study of the Bolshoi and Millennium simulations}

\author[S.~Contreras et al.]
  {
S.~Contreras$^{1}$,
N.~Padilla$^{1,2}$,
C.D.P.~Lagos$^{3,4}$.
\\
 $^{1}$Instit\'uto Astrof\'{\i}sica, Pontifica Universidad Cat\'olica de Chile, Santiago, Chile.\\
 $^{2}$Centro de Astro-Ingenier\'{\i}a, Pontificia Universidad Cat\'olica de Chile, Santiago, Chile \\
 $^{3}$International Centre for Radio Astronomy Research (ICRAR), M468, University of Western Australia, 35 Stirling Hwy, Crawley, WA 6009, Australia.\\
 $^{4}$Australian Research Council Centre of Excellence for All-sky Astrophysics (CAASTRO), 44 Rosehill Street Redfern, NSW 2016, Australia.\\
}

\def\LaTeX{L\kern-.36em\raise.3ex\hbox{a}\kern-.15em
    T\kern-.1667em\lower.7ex\hbox{E}\kern-.125emX}

\begin{document}

\pagerange{\pageref{firstpage}--\pageref{lastpage}} \pubyear{2016}

\maketitle

\label{firstpage}
\begin{abstract}
We use three different cosmological dark matter simulations to study how the orientation of the angular momentum vector (AM) 
in dark matter haloes evolve with time. We find that haloes in this kind of simulations are constantly affected by a spurious 
change of mass, which translates into an artificial change in the orientation of the AM. After removing the haloes affected by 
artificial mass change, we found that the change in the orientation of the AM vector is correlated with time. The change in its 
angle and direction (i.e. the angle subtended by the AM vector in two consecutive timesteps) that affect the AM vector has a dependence 
on the change of mass that affects a halo, the time elapsed in which the change of mass occurs and the halo mass. We create a Monte-Carlo 
simulation that reproduces the change of angle and direction of the AM vector. We reproduce the angular separation of the AM vector since a 
look back time of 8.5 Gyrs to today ( $\rm \alpha$) with an accuracy of approximately 0.05  in $\rm cos(\alpha)$. We are releasing this Monte-Carlo 
simulation together with this publication. We also create a Monte Carlo simulation that reproduces the change of the AM modulus. We find that haloes 
in denser environments display the most dramatic evolution in their AM direction, as well as haloes with a lower specific AM modulus. These 
relations could be used to improve the way we follow the AM vector in low-resolution simulations.

\end{abstract} 


\section{Introduction} 

Semi-analytic models (SAMs) are an efficient and accurate way to populate large volume simulations with galaxies 
\citep{Lagos:2008,Lagos:2012,Baugh:2006}, but in most cases, this hampers their ability to resolve halo and galaxy 
properties that are important for small-scale physical processes, such as star formation and feedback.  For example, 
some models relate the star formation rate and the mass loading of stellar-driven winds to the density of gas in the 
galaxy disc \citep{Croton:2006,Lagos:2013}, which critically depend on the ability of models to compute the 
sizes of galaxies. Typically models do the latter by assuming conservation of angular momentum (AM), from which a good 
measurement of the AM of dark matter haloes is crucial (e.g. \citealt{Cole:2000}).  This is usually done following Mo, Mao \& White (1998, MMW)  \nocite{Mo:1998}
who used detailed numerical simulations of the process of gas cooling, to find a simple analytic relation between the 
sizes of galaxies and the specific AM (sAM) of halos that could be used by simpler models. Thus, this relation implies 
that we need to be able to measure the AM of haloes, which cannot be accurately performed for structures traced by less 
than $\sim 1000$ particles \citep{Bett:2010}.  Most SAMs use merger trees that use as few as $20$ particles to define 
dark matter halos (e.g. the Millennium and Millennium-II simulations; \citealt{Springel:2005,BK:2009}), 
especially in large volume simulations.

There are two possible solutions for this problem. The simpler one is to use a sufficiently high-resolution simulation 
to limit the SAM to populate haloes with at least $1000$ particles, as it is done by \cite{Guo:2011}.  But in large 
volume simulations, this would imply very poor sampling of the dark matter halo mass function, with the smallest haloes 
having masses of $\approx 10^{12}\,\rm M_{\odot}$ (e.g. in the Bolshoi simulation, \citealt{Klypin:2011}).  
Another solution is to use Monte-Carlo simulations to follow the evolution of the AM of haloes, which in turn can be used to 
obtain the AM of galaxy discs using MMW.  This approach was adopted by Padilla et al. (2014, see also \citealt{Lagos:2014,Lagos:2015a}), \nocite{Padilla:2014}
where a SAM uses the merger trees of haloes defined with as few as  $10$ particles, but does not attempt to use the measured halo AM. 
It rather adopts a Monte-Carlo approximation relating changes in the AM to changes in the mass of the halo, also separating halo mass 
accretion between smooth accretion and mergers. This approximation provides an accurate prediction of the relative orientations of gas 
and stellar discs in observations of early-type galaxies \citep{Lagos:2014,Lagos:2015a}.

In this work, we expand the analysis of the evolution of AM of haloes including not only the change in direction of the AM vector 
of haloes, but also the relation between changes taking place at different times throughout the life of a halo.  We provide relations 
that can be used to improve the modelling of the AM of haloes in simplified models of galaxy formation, including SAMs, but also applicable 
to the subhalo abundance matching (SHAM,  \citealt{Kravtsov:2004,Vale:2006,Conroy:2006}) or the 
halo occupation distribution (HOD, \citealt{Jing:1998a,Peacock:2000}) type modelling.

\label{Intro}
\section{The dark matter simulations}

In this work we use three different dark matter simulations. The Bolshoi simulation \citep{Klypin:2011}, 
the Millennium simulation (\citealt{Springel:2005}, hereafter MS-I) and the Millennium-II simulation (\citealt{BK:2009}, hereafter MS-II). 
The properties of these simulations are listed in table 1. The Bolshoi simulation was run with an Adaptive-Refinement-Tree (ART) code, which is 
an Adaptive-Mesh-Refinement (AMR) type code (see \citealt{Kravtsov:1997} and \citealt{Kravtsov:1999} for a description of the code), while the MS-I and the 
MS-II were run with GADGET-2 and GADGET-3 \citep{Springel:2001b,Springel:2005}, respectively, and a TreePM code.

The Bolshoi simulation adopted a WMAP-5 cosmology ($\Omega_{\rm b}$ =0.045, $\Omega_{\rm M}$ = 0.27, $\Omega_{\Lambda}$ = 0.73, 
h = $H_0/100$ = 0.7, $n_{\rm s}$ = 0.95, $\sigma_8$ = 0.82) while the Millennium simulations adopted  a WMAP-1 based cosmology 
$\Omega_{\rm b}$ =0.045, $\Omega_{\rm M}$ = 0.25, $\Omega_{\Lambda}$ = 0.75, 
h = $H_0/100$ = 0.73, $n_{\rm s}$ = 1, $\sigma_8$ = 0.9). The difference in cosmology between the models should not cause 
significant differences in the results we present here.

The most relevant difference between the halo catalogues from these simulations is the halo finder algorithm and the merger 
trees used. The Millennium simulations use a Friends-of-Friends (FoF) group-finding algorithm \citep{Davis:1985}. The haloes 
are identified in each simulation output (snapshot) and contain a minimum of 20 particles.  The halos in consecutive snapshots 
are connected to build merger trees. The Bolshoi simulation uses the Rockstar halo finder algorithm \citep{Behroozi:2013a} 
and Consistent Trees \citep{Behroozi:2013b} to build merger trees. The latter codes are thought to improve both the completeness 
(through detecting and inserting otherwise missing halos) and purity (through detecting and removing spurious objects) of both 
merger trees and halo catalogues \citep{Behroozi:2013b}. These latter qualities are desirable for studies of the evolution of 
the AM vector.

Another important difference between the two halo catalogues is the amount of snapshots available and length of the timesteps. 
The MS-I and MS-II have 63 and 67 snapshots available, respectively, between z=127 to z=0. The length of the timesteps varies to 
allow a better time resolution at low redshifts. The Bolshoi simulation has 181 snapshots from z=14 to z=0. In Bolshoi, these 
snapshots are normally separated by $\Delta a = 0.03$ (the expansion factor) between $a=1$ to $a=0.8$ and  $\Delta a = 0.06$ at 
earlier snapshots for most cases.

The haloes from the Millenium simulations were obtained from the German Astrophysical Virtual Observatory (GAVO)\footnote{\url{http://gavo.mpa-garching.mpg.de/portal/}}, 
while the haloes from the Bolshoi simulation were obtained from Behroozi's personal webpage\footnote{\url{http://www.slac.stanford.edu/~behroozi/Bolshoi_Catalogs/}}.

\begin{table}
\begin{center}
    \begin{tabular}{| l | l | l | c |}
    \hline
    Simulation & $ N_{\rm P}$ & $ m_{\rm P}/h^{-1} {\rm M_{\odot}}$ & $ L/ h^{-1}{\rm Mpc}$\\ \hline
    Bolshoi & $2048^3$ & $1.4 \times 10^8$& 250 \\ \hline
    MS-I & $2160^3$ & $8.61 \times 10^8$& 500 \\ \hline
    MS-II & $2160^3$ & $6.88 \times 10^6$& 100 \\ \hline
    \hline
    \end{tabular}
\caption{Parameters of the Bolshoi, Millennium and Millennium II dark matter simulations. $N_{\rm P}$ represent the number of 
particle for each simulation, $ m_{\rm P}$ the mass of these particles and L is the length of the simulations' periodic boxes.  } 
\end{center}
\end{table}

\label{Sec1}
\section{Halo angular momentum and selection}

The AM vector of a dark matter halo is computed summing the cross product of the dark matter particles position with respect the halo center of mass ($\vec{r} - \vec{r}_{COM}$) 
and their momentum vector with respect to the halo center of mass ( $m(\vec{v}-\vec{v}_{COM})$, where $m$ is the mass of the particle),
\begin{equation}
\vec{J} = \sum_i m_{i}\ (\vec{r}_i - \vec{r}_{COM}) \times (\vec{v}_i-\vec{v}_{COM}).
\end{equation}

To ensure an accurate measurement, we impose a minimum number of particles per halo of 1000 particles. 
Whenever we follow a halo through different snapshots, we require the halo to satisfy this limit at all redshifts, 
effectively increasing the final, low redshift halo mass of samples of haloes followed through longer periods of time.
A less conservative value for the minimum number of particles required to robustly measure the angular momentum 
of dark matter haloes has been recently proposed by \citealt{Benson:2017} 
(i.e., 40,000 particles). To show the effect of a higher cut in the number of particles, most of the plots in this work will include the 
results for halo samples with different number of particles.

Before a halo undergoes a merger, and when it is not the largest halo in the merger (i.e. it is not part of the main 
progenitor branch of its descendants) or when the halo suffers a flyby, it loses a considerable number of dark matter 
particles. These losses, which can be driven by numerical errors due to the halo finder rather than to an actual mass 
loss, can seriously affect the AM direction of the halo.  In several cases, this mass loss is followed by a mass gain 
as the halo orbits away from the main halo \citep{Jiang:2014}.  This mass gain is also not of a physical origin.  
From this point on, we will refer to this type of mass variations as spurious mass changes. In this work, we are only 
interested in the change of angle of the AM caused by smooth accretion of individual dark matter particles, or by mergers 
with smaller haloes. To ensure this we only use haloes that are part of the main progenitor branch and have a descendant at 
z=0. We acknowledge that halo mass can decrease due to different effects such as tidal stripping and flybys, and these 
effects will also be included in our analyses.

\label{Sec2}
\subsection{Defining the cleaned halo samples}

\begin{figure}
\includegraphics[width=0.45\textwidth]{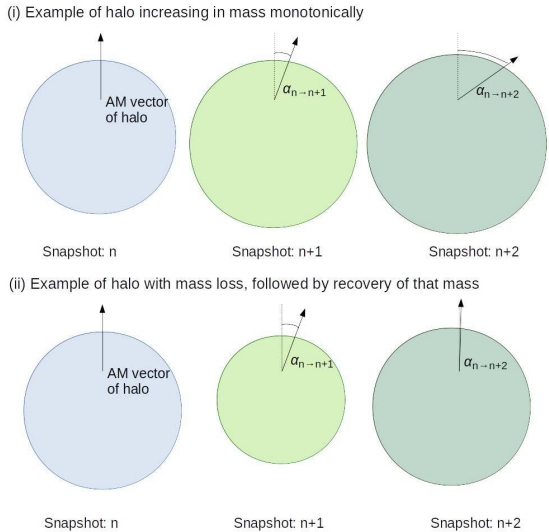}
\caption{Schematic showing examples of the change in direction of the AM of haloes due to real mass increase, for example, due to 
smooth mass accretion (top panel), and spurious loss of mass for a snapshot, that is later recovered by the halo (bottom panel). 
The angles $\alpha_{\rm n\rightarrow n+1}$ and $\alpha_{\rm n\rightarrow n+2}$ are shown in the middle and right hand circles. Here 
each circle represents a DM halo at a given time and its size represents mass, while the solid arrows show their AM vector. The dotted 
lines in the middle and right hand circles represent the AM vector at the snapshot $n$. }
\label{Fig:1}
\end{figure}

In this section we study the change of the AM of haloes in pairs of consecutive snapshots, n, n+1.  We use the angle subtended 
by the AM in the snapshots n and n+1, $\Delta \alpha$, as a way to detect spurious changes.  If the particles a halo appears to 
lose are recovered later, the measured AM will show a sudden change at the moment of particle loss, to then go back practically 
all the way to its previous value once the particles are regained by the halo. In the top panel of Fig.~\ref{Fig:1} we show schematics of 
what we would expect for a real change in the direction of the AM vector in dark matter halos due to a systematic change in mass, 
while in the bottom panel of Fig.~\ref{Fig:1} the schematic shows the effect of spurious mass loss for a timestep, that is later recovered, 
causing the AM vector to return to its original value.

\begin{figure*}
\includegraphics[width=0.90\textwidth]{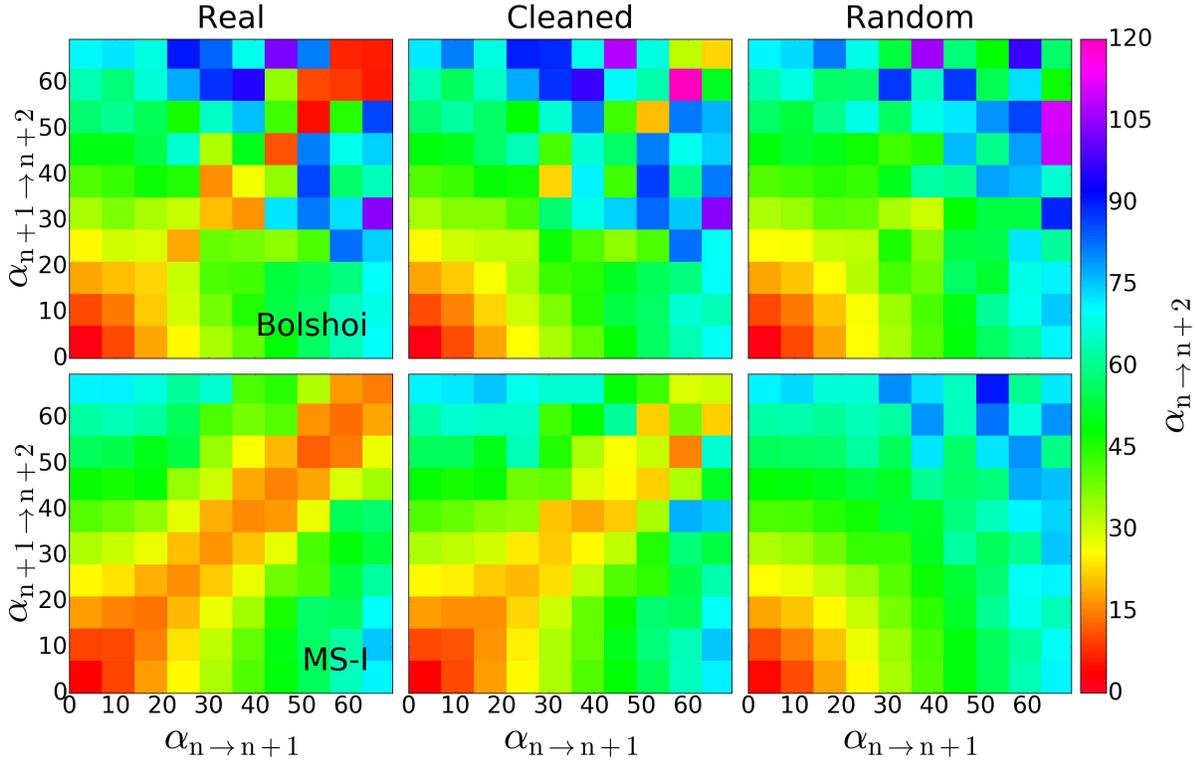}
\caption{Left panels: The angle between the AM of halos in consecutive snapshots $n+1$ and $n+2$, as a function of the angle between 
the AM of halos in snapshots $n$ and $n+1$. Pixels are coloured by the median angle subtended by the change of the AM vector suffered by haloes 
between a fixed snapshot ``n'' and their descendants in two future snapshots  ($\alpha_{n\rightarrow n+2}$, see the colour bar for the scale), for 
the Bolshoi (top panels) and the Millennium (bottom panels) simulations. The $n+2$ snapshots shown correspond to z=0. Middle panels: same as left 
panels, but taking out all haloes that in snapshot $n+2$ recover mass they lost in the previous snapshot (possibly just noise in the total halo mass 
rather than actual mass change). Right panels: same as the left panels, but for a Monte Carlo halo sample forced to have the same change of angle of 
their AM vector as the real haloes; each change of angle is randomly oriented with respect to the previous change of angle.}
\label{Fig:2}
\end{figure*}

Fig.~\ref{Fig:2} shows the effect of spurious mass changes in the change in direction of the AM of halos.  The left panel shows the angle subtended by 
$\vec{J}$ between two non-consecutive snapshots (leaving one snapshot in between).  We refer to this angle as $\alpha_{n \rightarrow n+2}$.  We show its 
value as a function of the angle subtended by the AM in the two pairs of consecutive snapshots that result from using the intermediate one, $\alpha_{n \rightarrow n+1}$ 
and $\alpha_{n+1 \rightarrow n+2}$.  The results are shown for the Bolshoi and the MS-I simulations (top and bottom panels). The results for MS-II look very similar to those in MS-I.

The left panels show very small  values of the angle subtended by the AM vector at $n$ and $n+2$, showing that the large variations between snapshots 
$n$ and $n+1$ are cancelled out  by those between $n+1$ and $n+2$, indicating a spurious change of the AM vector, which is not as apparent once more than 
two consecutive snapshots are examined.

This is the case for all the simulations studied here (including MS-II not shown in the figure) and all halo mass ranges. We find that the haloes that 
suffer from this effect tend to lose mass and then recover it back again usually after one or two snapshots.  However, we notice that there are extreme 
cases where haloes can lose and recover mass several snapshots later. 

This loss and later recovery of matter are caused when the halo finder algorithm stops associating a portion of dark matter particles as part of the main halo, 
a problem that can be present throughout several snapshots. The inverse process (an introduction and later extraction of particles) is also common and also produces 
this effect. This can be caused by flybys, tidal stripping, and virialization. As was mentioned above, the ROCKSTAR algorithm tries to avoid this effect by looking at 
a number of snapshots to the past and future to avoid including (or removing) particles that do not (do) belong to the halo.  Indeed, haloes in the Bolshoi simulation 
(where the ROCKSTAR algorithm was applied) are less affected by this problem.
 
To avoid the spurious effects driven by these haloes in our study, we create a halo sample where haloes that have suffered a significant change in mass and then are seen 
to recover it (or lose it again) are removed. We refer to this sample as the ``clean halo sample''. To make it to this sample haloes must satisfy the relation,
\begin{equation}
 |\Delta M_{n\rightarrow n+1}+\Delta M_{n+1\rightarrow n+2}| > C( |\Delta M_{n\rightarrow n+1}| + |\Delta M_{n+1\rightarrow n+2}|),
\end{equation}
\noindent
where $C$ is constant set equal to $0.1$ in the Bolshoi simulation and to $0.3$ in MS-I.  If $\Delta M_{n\rightarrow n+1} \sim - \Delta M_{n+1\rightarrow n+2}$, we 
remove the halo from the sample. To avoid removing haloes that had no accretion, and whose change in mass is only due to numerical noise in the halo finder algorithm 
we use the condition that both, $\Delta M_{n\rightarrow n+1}/M_{n+1}$ and $\Delta M_{n+1\rightarrow n+2}/M_{n+2}$ are below $10^{-3}$ to include a halo in the sample.  
These values are selected so as to minimise the impact of the rejected haloes in the main results presented throughout this paper.

\begin{figure}
\includegraphics[width=0.45\textwidth]{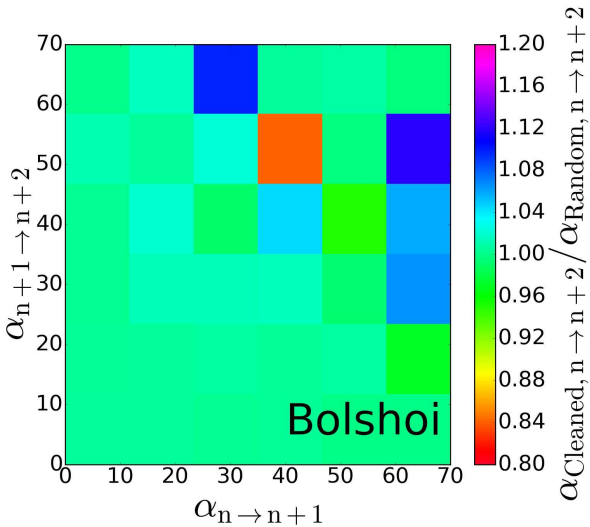}
\caption{ The ratio between the top middle panel of Fig.~\ref{Fig:2} ($\alpha_{n\rightarrow n+2}$ for the cleaned halo sample) with the top right panel of Fig.~\ref{Fig:2} 
($\alpha_{n\rightarrow n+2}$ for the random halo sample).}
\label{Fig:3}
\end{figure}

The middle panels of Fig.~\ref{Fig:2} show the results for the cleaned halo sample. We notice that for both simulations, the angles of the AM vector change in the diagonal are 
now larger than in the left panels, effect this is particularly strong for the Bolshoi simulation. This can be compared to the right panels of the figure where we 
choose a random orientation for the change of the AM vectors of haloes (i.e. with no coherence in the direction in which the AM vectors are moving towards between 
consecutive snapshots).  The similarity between the middle and right panels indicates that most of the changes of AM in the haloes appear to be mostly uncorrelated. 
However, as can be seen by in Fig.~\ref{Fig:3}, where we compare the middle and right panels of Fig.~\ref{Fig:2}, there are deviations between the two, as the angle between angular 
momenta of snapshots $n$ and $n+2$ is slightly smaller in the measured case than when the direction is randomly chosen. This indicates that the change of direction 
could be non-Markovian.  We present a  more detailed analysis in the following sections to quantify this more accurately.

\label{Sec3p1}
\section{Statistics of AM vector change in dark-matter only haloes}

In this section, we search for statistics of the rate of change of AM of haloes both, in their amplitude and direction.  Our aim is to produce a set of lookup tables 
that can then be adopted in simple models of galaxy formation such as SAMs, to include a physical evolution of the AM of galaxy host halos that is also numerically robust.  
We study the change of direction of the AM in two steps.  First we will concentrate on the angle subtended by the AM of a halo in pairs of snapshots.  Then we will study how 
correlated the changes in the AM are throughout the lives of DM haloes.

\label{Sec4}

\subsection{Evolution  in the direction of the AM vector}
\begin{figure}
\includegraphics[width=0.45\textwidth]{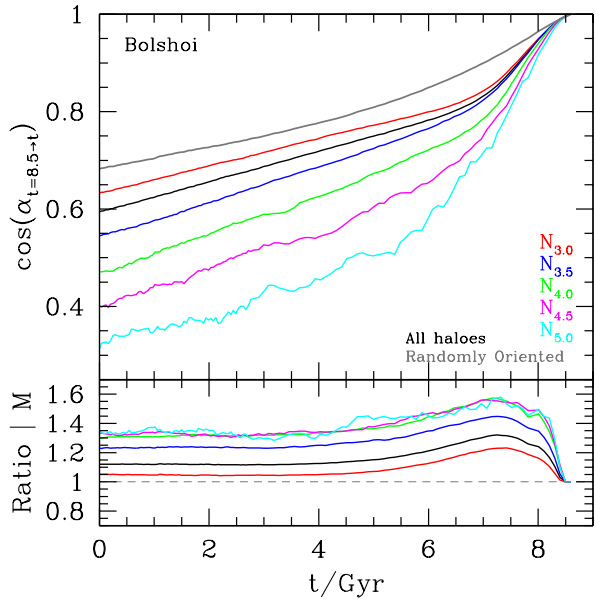}
\caption{ 
The top panel shows the median change of direction of the AM vector suffered by haloes between a lookback time t=8.5 Gyrs and their descendants at a lookback time ``t'' 
in the Bolshoi simulation. The black line represents the median of the full halo population. The colour lines represent different numbers of particles per halo, N3, N3.5, N4, N4.5 and N5, 
corresponding to $10^3-10^{3.5}$, $10^{3.5}-10^4$, $10^4-10^{4.5}$, $10^{4.5}-10^5$ and more than $10^5$ particles respectively at lookback time t=8.5 Gyrs. The grey line represents the median 
change of direction of a Monte Carlo generated halo evolution which uses the actual measured changes of direction as the real halo sample, but randomly oriented (as the right panels of Fig.~\ref{Fig:1}). 
As a reference, the 20th and 80th percentile of the distribution for the black line (all haloes) have a value of 0.08 and 0.94 at t=4Gyrs, and -0.11 and 0.9 at t=0Gyrs. Bottom panel: the 
ratio between the measured direction changes and the Monte Carlo result (random orientations).}
\label{Fig:4}
\end{figure}

The top panel of Fig.~\ref{Fig:4} shows the median cosine of the angle subtended by  $\vec{J}$ measured at a lookback time of 8.5 Gyrs and at a later lookback times ``t''.  Coloured lines represent a 
different number of particles per halo as indicated in the figure description. The black line represents the full galaxy population. This choice of line colours will be used in several other 
figures of this paper. As can be seen, the most massive haloes tend to deviate more from its original AM compared to the less massive samples. This measurement is not affected by the spurious 
mass changes discussed in the last chapter since we measure difference among several snapshots and not consecutive ones.

The grey line shows the predictions of the random orientations case. In this case, the angular separation is lower than measured in the simulation. 
This confirms that the change in the direction of $\vec{J}$ is correlated in time, and not randomly oriented.  The case of random orientations 
for low and high halo masses have a similar behaviour (not shown in the figure). The ratio between the median angular 
separation of haloes and their counterpart of the random case is shown in the bottom subpanel of Fig.~\ref{Fig:4}.

\begin{figure}
\includegraphics[width=0.45\textwidth]{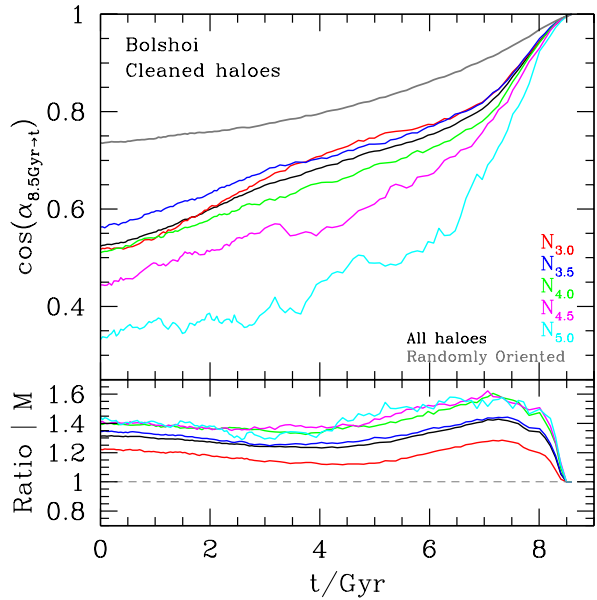}
\caption{ 
Same as Fig.~\ref{Fig:4} but for a selection of haloes that are not affected artificial mass changes (cleaned halo sample, see section~\ref{Sec3p1} for more details).}
\label{Fig:5}
\end{figure}

Fig.~\ref{Fig:5} is similar to Fig.~\ref{Fig:4}, except that we only show haloes with no spurious mass changes in any of the $121$ snapshots at lookback time $8.5$Gyr or later 
in the Bolshoi simulation. The latter is a very restrictive selection and thus we remove most of the halos of the simulation, but the remaining haloes represent a clean sample 
with little numerical artefacts. We notice that the random case shows departures from the initial direction of the AM that are larger than in Fig.~\ref{Fig:4}. This is due to the lower 
amount of false angle changes. These false changes of angle do not affect the total change of angle (as is shown in Fig.~\ref{Fig:5}) but affect the mock random sample by increasing total 
angle separation in time. The results for haloes of different masses show a larger angular separation than in the case of using all haloes (Fig.~\ref{Fig:4}). Even when adopting different 
parameters for Eq. 2, and also trying different recipes to clean spurious mass changes in the sample, the result remains qualitatively similar.  Cleaning the cases of spurious mass 
changes affect primarily the results for the low mass haloes, making them depart more strongly from their initial AM direction.

\begin{figure}
\includegraphics[width=0.45\textwidth]{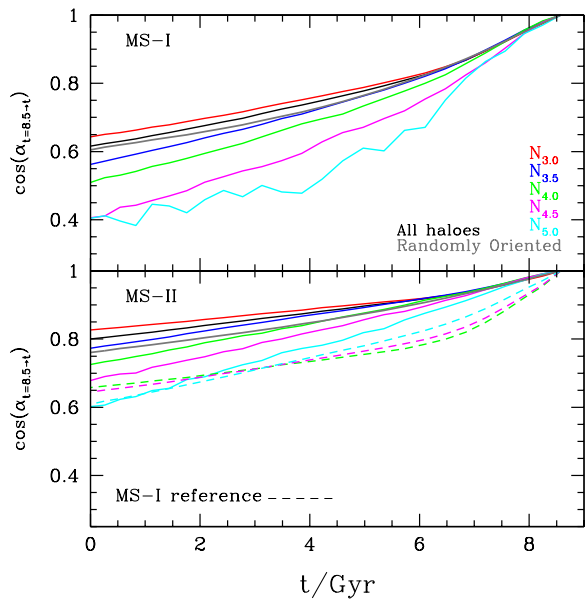}
\caption{
Same as the main panel of Fig.~\ref{Fig:2} but for the MS-I (Top) and MS-II (Bottom).  The dashed lines in the bottom panel show the MS-I results for the same mass 
ranges in the three most massive bins in MS-II, for comparison.
}
\label{Fig:6}
\end{figure}
 
 \begin{figure*}
\includegraphics[width=1.0\textwidth]{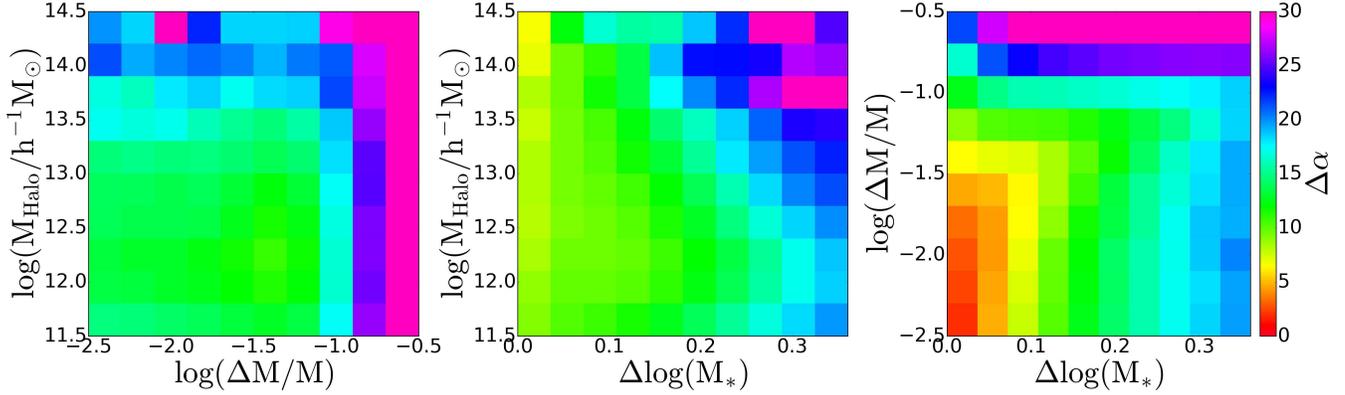}
\caption{The median change of direction of the AM vector as a function of three factors: the halo mass ($M_{Halo}$), the difference in $M_*$ ($\Delta log(M_*)$) and 
the change of mass suffered by the halo between two redshifts ($\Delta M/M$). The left panels show the dependence with respect to  $M_{Halo}$ and $\Delta M/M$, the 
middle panels show the dependence with  $M_{Halo}$ and $\Delta log(M_*)$, and the right panels show the dependence with respect to $\Delta M/M$ and $\Delta log(M_*)$. 
Whenever we compare two different properties, we select haloes with a value of the third halo property as close as possible to its median value, except for $\Delta log(M_*)$ 
where we select haloes with a value between 0.2 and 0.3. Haloes from all redshift ranges are used for these plots.
}
\label{Fig:7}
\end{figure*}

Fig.~\ref{Fig:6} shows the median change of angle of $\vec{J}$  for the MS-I (top) and the MS-II (bottom).  For both simulations, the random case (presented 
also in grey line) shows a larger angular separation than measured for low mass haloes (and for the full sample). This could be easily misinterpreted as the 
change of angle of  $\vec{J}$  in these simulations being uncorrelated (or even anticorrelated). But the cause behind this result is the spurious mass change 
of haloes due to numerical noise affecting the random sample.
Once the mass is recovered the AM mostly goes back to its original direction, an effect that is not present in the random case. This is the same effect that 
caused Fig.~\ref{Fig:5} (the angular separation of the cleaned sample of the Bolshoi simulation) to show larger differences than Fig.~\ref{Fig:4} (same as 
Fig.~\ref{Fig:5}, but for the full halo population).

The question remains on whether the change of direction of the AM vector is related to the mass of the haloes, or a numerical effect produced by the difference 
in a number of particles sampling the mass distribution in the haloes.  To answer this we also show in the bottom panel of Fig.~\ref{Fig:6} on dashed lines the haloes from 
the MS-I simulation, for ranges of particles that correspond to equal mass ranges of MS-II samples.  Since the volume of the MS-II is smaller, this can only be done 
for the three most massive samples.  To be able to do this comparison we use haloes with at least  $80$ particles. Also, the most massive sample includes MS-I haloes
with masses corresponding to $10^5$-$10^{5.5}$ particles per halo in the MS-II. There is no agreement between the behaviour of haloes of similar mass in the MS-I and 
MS-II simulations, regardless of the mass range studied.  In the case of the most massive sample shown in the bottom panel for MS-I, the minimum number of particles 
is high enough to obtain accurate measurements of AM direction (a total of $800$).

A possible reason for the difference is the environment, as while in MS-II the sample corresponds to the most extreme haloes, which probably live in the 
densest environments, in the MS-I simulation these haloes are actually the smallest ones which live in several different types of environments. This could 
indicate that mass is not the only property of the halo that is relevant for the evolution of $\vec{J}$, a subject we will come back to in Section~\ref{Sec6}.

Fig.~\ref{Fig:7} shows $\Delta \alpha$ as a function of combinations of halo mass ($M_{\rm Halo}$), the difference between the snapshots of the logarithm of 
the characteristic mass at redshift z ($\Delta log\ M_*(z) $) and the change of mass suffered by the halo in that time 
($\Delta M/M$, calculated as $(M_{\rm final}-M_{\rm initial})/M_{\rm final}$) for the Bolshoi simulation. We allow pairs of snapshots separated by up to $30$ 
snapshots to allow larger redshift differences. Whenever we show values of $\Delta \alpha$ as a function of any two different properties, we do so by fixing the 
remaining variable to it corresponds to the median value, except for $M_*$, where we select haloes with $\Delta log\ M_*(z) $ between 0.2 and 0.3 (this is 
an arbitrary cut; we test other values finding similar trends as the ones shown here).

We use $\Delta log\ M_* $ instead of elapsed time, since we find that there is less redshift dependence of the results, i.e. the evolution of two haloes that 
evolve in a constant $\Delta log\ M_* $ at different redshift are similar compared to the evolution of two haloes that evolve during the same time at different 
redshifts. We also test other variables like the redshift, the expansion factor($a$) the growth factor ($g$) the amplitude of the growing mode ($D$), but no 
property work as good as $M_*$. 

To calculate $M_*$, we follow a similar procedure of the one presented in \citealt{RP:2016} to calculate the characteristic mass of halos just collapsing
at redshift $z$ ($M_C$). This is shown in Appendix A.

As can be seen $\Delta \alpha$ shows a strong dependence on the range of $M_*$ and the mass variation, where a higher $\Delta \alpha$ takes place when the timestep is 
longer or the change of mass is larger, as expected. The halo mass shows little dependence with $\Delta \alpha$, where for small timesteps, high halo masses undergo a small
change of angle. For long timesteps the relation reverses. 
 This is consistent with the evolution seen in $\Delta \alpha$ and the dependence on halo mass in the case where random orientations are applied.

\label{Sec31}

\subsection{Persistence of angular momentum changes}

One of the aims of this paper is to find a way to follow the evolution of the AM vector in halo samples in low-resolution (or large volume) dark matter simulations. 
To do this it is not enough to reproduce the amplitude of the change of AM direction because, as we see with the mock random sample, this would correspond to the 
random case shown in Figures~\ref{Fig:4},~\ref{Fig:5} and~\ref{Fig:6}, and it does not reproduce the behaviour followed by the AM vector of the haloes in our simulations. In addition to the 
amplitude of the change, we also need to reproduce the persistence of the direction of change in order to reproduce the evolution of $\vec{J}$ in the long run.  
We need to quantify if when there is a change of AM between a given pair of snapshots, the following changes occur in a direction that is related to the previous one.  
That is, whether the projection of $\vec{j}_{m'}-\vec{j}_{n}$ and $\vec{j}_{m}-\vec{j}_{m'}$ over a plane perpendicular to $\vec{J}_{m'}$ of different pairs of 
snapshots $n<m'<m$ are parallel or not. We define the direction (DIR) parameter as follows:

\begin{equation}
cos(DIR) = \hat{\Delta J}_{1} \cdot  \hat{\Delta J}_{2},
\end{equation}

\begin{equation}
\Delta J_1 =  \hat{J_{n}}cos(\alpha_1)-\hat{J_{n-1}},
\end{equation}

\begin{equation}
\Delta J_2 =  \hat{J_{n+1}}-\hat{J_{n}}cos(\alpha_2),
\end{equation}
 \noindent
where $\alpha_1$ and $\alpha_2$ are the angle between $\hat{J_{n-1}}- \hat{J_{n}}$ and $\hat{J_{n}}- \hat{J_{n+1}}$, respectively.

 \begin{figure*}
\includegraphics[width=1.0\textwidth]{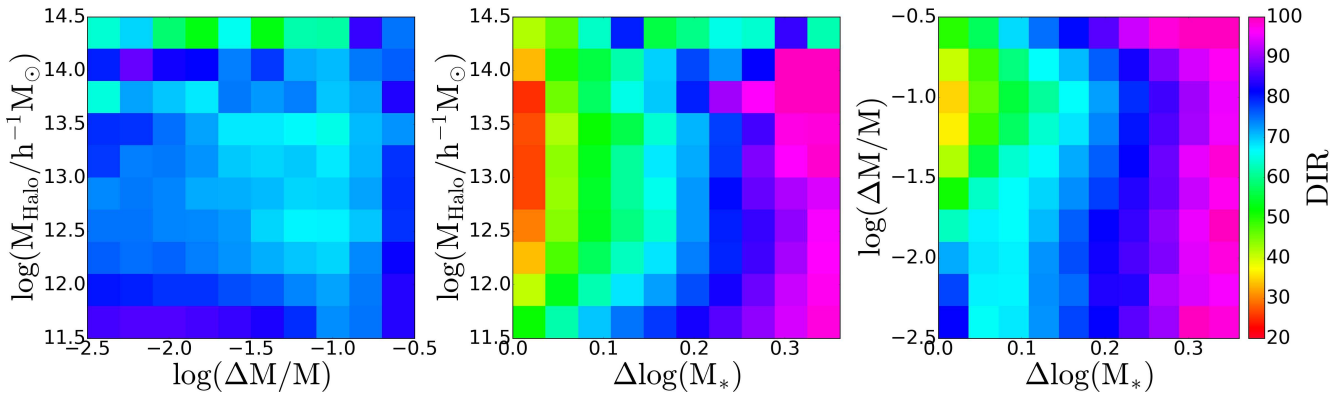}
\caption{The median change angle between consecutive direction changes of the AM vector as a function of three factors: the halo mass 
($M_{Halo}$), the difference in the logarithmic of $M_*$ ($\Delta log(M_*)$) and the change of mass suffered by the halo between two 
redshifts ($\Delta M/M$). Panels are as in Fig.~\ref{Fig:7}.
}
\label{Fig:8}
\end{figure*}

A value of $DIR = 0^o$ indicates that the change of AM continues in the same direction (if we have a change of angle of $3^o$ and $1^o$, 
the total change of angle will be $4^o$).  A value $DIR = 180^o$ indicates anticorrelation (if we have changes of $3^o$ and $1^o$, the total 
change of angle will be $2^o$). If the median $DIR$ is $90^o$, the
change of angles is random (as in our random case). 

Fig.~\ref{Fig:8} shows $DIR$ as a function of combinations of the halo mass, the change of mass, and the change of the logarithmic of 
$\Delta log(M_*)$, and as can be seen, it has a low dependence on the halo mass, where more massive haloes have smaller values of DIR 
(i.e. changes in angle are less correlated). The direction has also a strong dependence on $\Delta log(M_*)$ and $\Delta M/M$. Larger 
$\Delta z$ and lower values of $\Delta M/M$ imply larger DIR values, but the relation is not as clear as in Fig.~\ref{Fig:7}. 

\label{Sec32}
\subsubsection{Introducing persistence in a Monte-Carlo evolution of the angular momentum}

\begin{figure}
\includegraphics[width=0.45\textwidth]{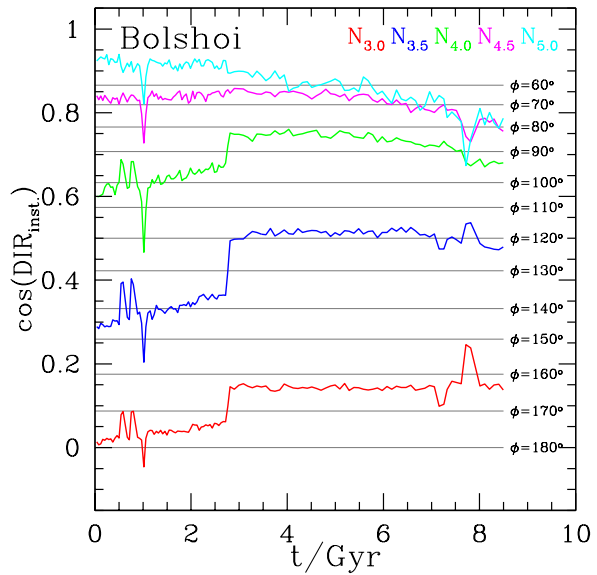}
\caption{The angle between two consecutive changes of AM direction (taken throughout three consecutive snapshots) as a function of the lookback time (t) 
for different halo mass ranges as labelled. The halo mass and the timestep between snapshots have a strong influence in the change of direction; this is
easily seen at $\sim 3Gyr$ (also in Fig. 7).  The grey lines show the predictions of the random samples with a banning angle ($\phi$) as labelled in the right 
part of the plot. By definition, the median direction of a mock random sample with a fixed banning angle will be not affected by the length of the timestep 
or the halo mass, and behaves as $\langle DIR_{inst.} \rangle = \phi/2$.
}
\label{Fig:9}
\end{figure}

A simple way to reproduce the median change of direction in a halo sample is to limit the allowed range of change of direction in the random case,
\begin{equation}
DIR <  2\langle DIR \rangle \equiv \phi,
\end{equation}
where $\phi$ is the maximum allowed angle of $\rm DIR$, which by construction follows a uniform distribution (the mean of a random and uniform sample, $\langle DIR \rangle$, is simply half the value of $\phi$). Fig.~\ref{Fig:9} shows the instantaneous 
change of direction of the AM vector in consecutive snapshots as a function of lookback time. The grey horizontal lines show the value corresponding to 
random changes with different values of $\phi$. Notice the jump in measured halo AM changes at t=3 Gyrs, which is caused by a change in the simulation timestep. 
As can be seen, the different halo mass ranges show roughly constant values of $DIR$ as a function of redshift, which indicates that the level of persistence in 
the direction of change of the AM vector is similar throughout the life of a halo, and much stronger for high mass objects.  We will use the measured changes of 
$DIR$ to interpolate a value of $\phi$ to apply to our random cases, in order to try and reproduce the observed evolution of the AM in a Monte-Carlo fashion.

We use the median change of angle and direction as a function of halo mass, interval of redshift and change in halo mass to attempt to reproduce the evolution of the AM vector.

\label{Sec321}

\section{Monte Carlo evolution of the halo angular momentum vector}
 \begin{figure}
\includegraphics[width=0.45\textwidth]{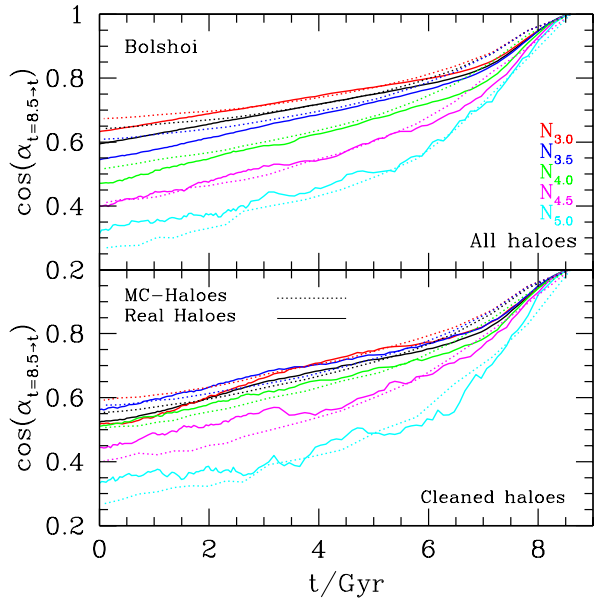}
\caption{(Top) Similar as the main panel of Fig.~\ref{Fig:4}. The predictions of the MC-Halo sample are shown as dotted lines. The change of angle and direction are 
calculated using the change of mass and redshift in consecutive snapshots. (Bottom) Same as the top panel, for but for the cleaned halo sample.
}
\label{Fig:10}
\end{figure}

We develop a Monte Carlo mock random sample (MC-Halo sample from this point on), to predict the change in the angle and direction of the AM vector of a dark matter halo, 
using the Bolshoi simulation. We later test the resulting model on the MS-I.

To create this Monte-Carlo simulation, we measure the median change of angle and direction that affect a halo in the Bolshoi simulation as a function of its mass, its change 
of mass in a given timestep and the change of $M_*$ in that snapshot. To avoid the contamination of haloes that suffer from spurious mass changes, we use the cleaned halo sample 
to create this simulation.

The predictions of the angular separation since a look back time of 8.5 Gyrs for the full halo sample and for the cleaned halo samples for the are shown in the top and bottom panel 
of Fig.~\ref{Fig:10}. To calculate the change of angle and the change of direction, we use time intervals in the Bolshoi simulation that are equal to 7 consecutive snapshots (similar to one MS-I 
snapshot time interval) to avoid the noise of the income and outcome of mass that typically affect haloes at short timescales. We manage to reproduce the evolution to 0.05 accuracy in 
$cos(\alpha)$ for most cases. We also test the performance of the MC-Halo sample run with tables that use all haloes (and not only the cleaned halo sample) finding similar results to the ones presented here.

\begin{figure}
\includegraphics[width=0.45\textwidth]{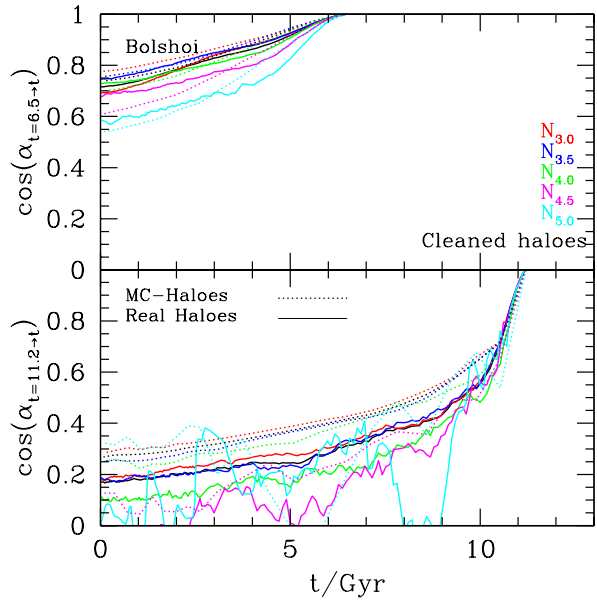}
\caption{ Similar as main to the bottom panel of Fig.~\ref{Fig:10}, but for an initial time of 6.5 (top) and 11.2 Gyrs (bottom).}
\label{Fig:10.1}
\end{figure}

\begin{figure}
\includegraphics[width=0.45\textwidth]{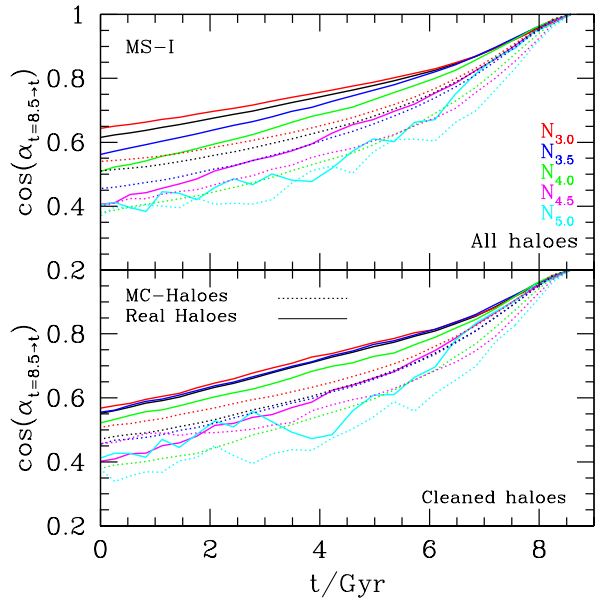}
\caption{ Similar as Fig.~\ref{Fig:10}, but for the MS-I}

\label{Fig:10.2}
\end{figure}

We test the performance of these tables by following galaxies at other initial times. The predictions of the angular separation since a look back time of 
6.5 and 11.2 Gyrs for the cleaned halo samples are shown in the top and bottom panel of Fig.~\ref{Fig:10.1}, respectively. We succeed at reproducing the evolution to 0.05 accuracy in $cos(\alpha)$ 
for an initial look back time of 6.5 Gyrs, and to 0.1 for an initial look back time of 11.2 Gyrs. Finally, in Fig.~\ref{Fig:10.2} we test our MC-simulation over the cleaned halo sample of the MS-I. 
The MC-Halo sample reproduces the same trends in the evolution of the angular separation of the real halo sample, but over predicts the angular separation by $\sim 0.1$ in $\cos{\alpha}$. 
This difference could be caused because the clean mechanism does not work properly on the MS-I, as it shows in Fig.~\ref{Fig:2}. A constant income and outcome of mass will artificially move the 
AM vector of the MC-Halo sample from its original position, making halos appear to have DIR closer to random than they truly have, which is consistent with what is shown here. Nevertheless, 
we find that a difference of 0.1 in $\cos{\alpha}$ in 8.5 Gyrs of evolution is acceptable considering that this is a different simulation from the one we used to create the tables, with a 
different volume, resolution and that the haloes are identified with a different halo finding algorithm. 

To provide a full evolutionary model of the AM vector of dark matter haloes, we also develop a Monte Carlo simulation that predicts the evolution of 
the modulus of the AM as a function of the same parameters used before. In addition, we use a modificated version of the relationship found by Catelan \& Theuns (1996, hereafter  
C\&T96) that links the modulus of the specific AM (sAM), $|j|$, with it halo mass. 
The model for the growth of the angular momentum of halos of C\&T96 has been tested against modern dark matter simulations (e.g. \citealt{Zavala:2008,Book:2011}) and for hydrodynamic simulations (eg. \citealt{Zavala:2016}), 
and proved to be a good approximation.
This relation assumes that, for a fixed redshift \nocite{Catelan:1996}
\begin{equation}
|j / h^{-1}Mpc\ km/s| = p\ (M/h^{-1}M_{\odot})^{q},
\end{equation}
\noindent
with $p$ and $q$ constants. We can add a temporal dependence to $p$ and $q$, 
\begin{equation}
|j / h^{-1}Mpc\ km/s| = p(a)\ (M/h^{-1}M_{\odot})^{q(a)},
\end{equation}
\noindent
with
\begin{equation}
log(p(a)) = 
 \left\{
    \begin{array}{ll}
        log(p_0) & \mbox{if } a > a_0\\
        log(p_1)\ (a-a_0)+ log(p_0)  & \mbox{if } a \leq a_0,
    \end{array}
\right.
\end{equation}
and
\begin{equation}
q(a) = 
 \left\{
    \begin{array}{ll}
        q_0 & \mbox{if } a > a_0\\
        q_1\ (a-a_0)+ q_0  & \mbox{if } a \leq a_0,
    \end{array}
\right.
\end{equation}

Here, $a$ is the expansion factor and $p_0$,  $p_1$,  $q_0$,  $q_1$ and $a_0$ are constants.
This is slightly different to what is proposed in C\&T 96, where they assume the value of $q$ constant, close to $2/3$ 
(expected power-law index in hierarchical cosmologies, MMW98). We fit the $|j|-M$ relation for different values of the scale 
factor, and we found that  $log(p_0)$,  $log(p_1)$,  $q_0$,  $q_1$ and $a_0$ are: -8.01, -0.77, 0.68, 0.09 and 0.71. 
This fit was done for values of $a$ between 0.3 and 1.  \cite{Liao:2015} also look into the evolution of $p$ and $q$ finding similar trends to the ones found by us.

\begin{figure}
\includegraphics[width=0.45\textwidth]{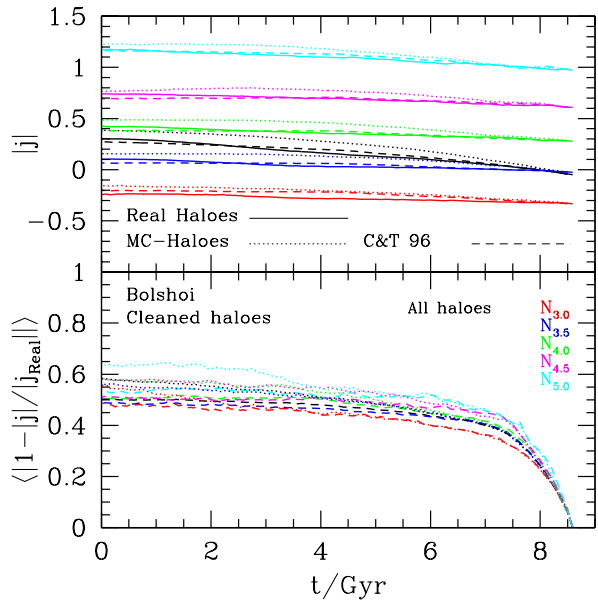}
\caption{The top panel shows the median value of the sAM for haloed sample between a lookback time t=8.5 Gyrs and their descendants at a lookback time ``t''. 
The solid line represent the cleaned halo sample of the Bolshoi simulation, while the dotted and dashed line show the prediction of the MC-Haloes and from 
the equation 7-11 (C\&T 96). The bottom panel show the average difference between the predicted modulus and the real modulus of the sAM vector. }

\label{Fig:10.3}
\end{figure}

To create the MC-Halo sample, with either the tables or using the analytic expression, we assume that:
\begin{equation}
|j(t + \Delta t,M + \Delta M)| = \frac{|j_{pred}(t + \Delta t,M + \Delta M)|}{|j_{pred}(t,M)|}|j(t,M)|,
\end{equation}
\noindent
where $j$ is the sAM of the halo and $j_{pred}$ the predicted sAM using the method explained above. By doing this, we can maintain 
the dispersion of the $|j|-M$ relation as we evolve the AM through time. This will also preserve most of the original spin distribution of haloes. The evolution of $|j|$ since a look back time of 8.5 Gyrs 
and the prediction of the MC-Halo sample using the tables and the modify expression of C\&T 96 is shown in the top panel of Fig.~\ref{Fig:10.3}.
We successfully reproduce the evolution of the median value of the sAM modulus of Bolshoi halos to an accuracy better than $10\%$ for 
the modified version of C\&T 96, and $20\%$ for the MC-Halo sample. When we compare the evolution obtained via these two methods with 
individual haloes (instead of the median growth), we notice that our prediction differs on average by $60\%$ ($50\%$) in 8.5 Gyrs of 
evolution for the MC-Halo (C\&T 96) samples, as is shown in the bottom panel of Fig.~\ref{Fig:10.3}. This is acceptable if we consider that both 
techniques follow the modulus of $|j|$ in a statistical way.

A public version of the tables used in all these calculations, and the code necessary to create an MC-Halo sample are now publicly available in \url{https://github.com/hantke/J_Tables}
\label{Sec5}
\section{Dependence with environment}

As was noted previously when analysing Fig.~\ref{Fig:6}, there is a strong difference in the evolution of the direction of the AM of haloes of equal mass in 
simulations with different resolution/volumes. The haloes in the MS-I showed a stronger change of direction than those in MS-II of similar mass. We 
pointed out that this could be due to the environment since in the MS-II such haloes correspond to the most massive ones, which probably lie in knots 
of the cosmic web, whereas in MS-I the haloes being compared can lie on more diverse environments.  

In this section we study this issue using the Bolshoi simulation, classifying the environment of haloes into voids, walls, filaments and knots, using 
the T-Web algorithm \citep{FR:2009,Hoffman:2012} publicly available in the CosmoSim virtual observatory\footnote{\url{www.cosmosim.org/}}. 
The T-Web is defined as the Hessian of the gravitational potential, and is computed on a $256^3$ cubic grid, which in the case of Bolshoi corresponds to cells 
with $\sim 1 h^{-1} Mpc$ a side. The tensor has three real eigenvalues in each grid. If 3, 2, 1 or 0 of those eigenvalues are above a given threshold, then the 
grid is classified as a knot, filament, wall or void, respectively. We choose $0.6$ as the value of the threshold. The T-Web was calculated only at z=0, the 
environment of the haloes was assigned depending on the location of haloes at z=0.  

\begin{figure}
\includegraphics[width=0.45\textwidth]{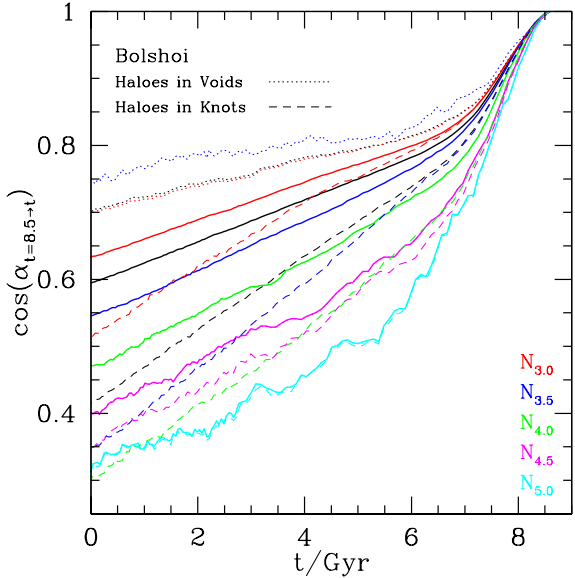}
\caption{Similar to the main panel of Fig.~\ref{Fig:4}. Haloes that at z=0 are located in knots and voids are shown in dotted and dashed lines respectively. 
The definition of knots and voids are done using the T-Web algorithm as is explained in section ~\ref{Sec6}.
}
\label{Fig:11}
\end{figure}

Fig.~\ref{Fig:11} shows the evolution of the direction of the AM of haloes since lookback time of 8.5 Gyrs for haloes that at $z=0$ lie in voids and knots. We find that 
haloes in knots (dashed line) tend to show larger deviations from their starting point, i.e. more persistent AM changes, than the average (solid line), and even 
more so compared to haloes living in voids (dotted line). 

We notice that there are no haloes in the three most massive bins that reside in voids, which is to be expected in underdense environments. Haloes in filaments 
and walls show an intermediate behaviour to those in void and knots, the two more extreme kinds of environments. The evolution of the AM is therefore strongly 
dependent on environment.

A more detailed study of the influence of the environment in the evolution of the AM vector in dark matter simulations is being done by Forero-Romero et al. in prep. 
In this paper, the authors will study how the direction and modulus of $\vec{J}$ are affected by the different definition of environment available in the literature.


\label{Sec6}

\section{Other dependencies and systematics}

\begin{figure}
\includegraphics[width=0.45\textwidth]{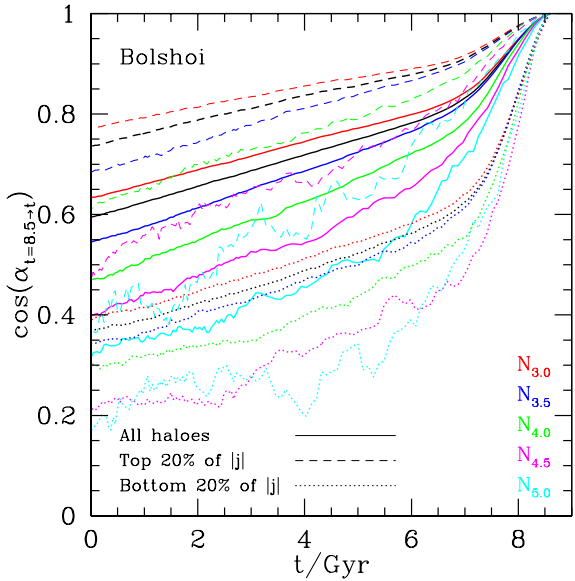}
\caption{
Similar to the main panel of Fig.~\ref{Fig:4}. Halos that have the 20\% lowest and higher AM modulus are shown in dotted and dashed lines respectively.
}
\label{Fig:12}
\end{figure}

Apart from the environment, we are interested in revealing other dependencies on intrinsic properties, 
such as the amplitude of the AM. Fig.~\ref{Fig:12} shows the angular separation of haloes between a lookback time of 8.5 
Gyrs and their descendant at a time ``t'' for haloes in the lower and higher $20$ percentiles of $|\vec{J}|$  
(shown in dotted and dashed lines, respectively). The angular separation of $\vec{J}$ between these two snapshots 
shows a strong dependence on its modulus. We find differences between these two bins of $|\vec{J}|$ of around 0.4 in $cos(\alpha)$ 
regardless of halo mass. This dependence is also valid for the sAM as we are comparing haloes of equal mass. Haloes with lower AM show a 
higher angular separation with their initial position compared to the haloes with a higher AM. Along with the previous tables of $\alpha$ 
and the direction as function of the halo mass, the redshift interval and the change of halo mass, we publish $\alpha$ and the direction as 
a function of these properties plus the modulus of the sAM so that this can be used in the generation of mock halo samples using the spin 
distribution of the dark matter haloes \citep{Gardner:2001}.

\label{Sec7}
\section*{Summary and Conclusions}

We studied how the orientation and the modulus of the AM vector evolve in time. We use three dark matter simulations (the Bolshoi simulation, the MS-I and the MS-II) and select haloes with a minimum of 1000 particles to have a good measurement of the AM vector. We found that there exist physical and numerical factors that affect the evolution of the AM vector. Here we summarise the main conclusions of our work:

\begin{itemize}
 \item Haloes are affected by spurious changes of mass. This is present in all the simulations studied here, but strongly on the 
 MS-I and the MS-II (where haloes were identified using solely the position distribution of particles). These changes of mass cause artificial slews on the AM vector of haloes. By cleaning the haloes affected by spurious mass changes (using Eq. 2), we also remove from the sample many low mass haloes that have a more passive evolution of their AM vector direction. We test other ways of cleaning these haloes, but in all cases, the resulting effect is the removal of haloes that have a passive evolution. These effects (the spurious mass changes and the biased cleaning method) should be taken into account before using the AM direction coming from any dark matter simulation.

 \item We measure the angle separation that affects an AM vector in time. We found that this angle separation is larger than the one obtained from uncorrelated direction changes in consecutive snapshots (i.e. random walks; we refer to this as the mock random sample). This means that the change in the direction of the AM vector is correlated in time. We also found that more massive haloes have a stronger correlation in the evolution of their AM direction.

 \item The change of angle that affects the AM vector has a strong dependence on the mass change suffered by the halo, as well as the time interval in which this change of angle occurs (expressed by the change in $M_*$). The angle of change of the AM vector displays a weak dependence on halo mass.

 \item The change of direction that affects the AM vector has a strong dependence on the change of mass that affects the halo, the change of  $M_*$ in which this change of angle occurs and also the halo mass. 

 \item Using the dependence mentioned above and the Bolshoi simulation, we create a Monte Carlo mock random sample (MC-Halo sample), capable of inferring the change of angle and the change of direction of the AM vector. With this simulation, we reproduce the angular separation since a look back time 
 of 8.5 Gyrs to today with an accuracy of  0.05  in $cos(\alpha)$ for most cases. {The performance of this MC-Halo sample is much better than the case in which we follow the change of angle between consecutive snapshots in an exact way, but without making use of the information of the change of orientation (equivalent to the randomly oriented case of Fig.~\ref{Fig:4}, Fig.~\ref{Fig:5}, Fig.~\ref{Fig:6}). In this case one underpredicts the total evolution of the orientation of the angular momentum and and can hardly distinguish between the evolution of haloes of different masses}.
 
 \item We test our MC-Halo sample on the MS-I simulation. We were able to reproduce the general behaviour of the evolution of the angular momentum vector with an accuracy of  $\sim$ 0.1  in $cos(\alpha)$. These predictions are not as good as for the Bolshoi simulation, but this is due to the merger trees, halo mass range, resolution and halo finder algorithm applied to this simulation being different to the Bolshoi simulation, the one used to make the MC look-up tables. However, the predicted evolution of the vector is still very well reproduced, and we consider it to be a better alternative to using the angular momentum vector measured with a small number of particles (as it has been traditionally done in semi-analytic models). 

 \item We also created a MC-Halo sample that reproduces the evolution of the modulus of the SAM of haloes, along with an analytic expression based on the work of C\&T96. We reproduce the modulus of $\vec{j}$ since a look back time of 8.5 Gyrs to the present day with a precision of more than a 90\% for the cleaned halo sample.

 \item The environment has a strong role in the evolution of the AM vector direction. Haloes in denser environment show a wilder evolution in the direction
 of their AM vector ($cos(\alpha) = 0.42$ in a 8.5 Gyrs evolution), compared to the median angle separation ($cos(\alpha) = 0.59$) and the 
 change suffered by halos in low density environments ($cos(\alpha) = 0.7$). The latter holds even at fixed halo mass.
 
 \item The evolution of the AM vector direction is also strongly correlated with the amplitude of the AM vector. Haloes with higher AM modulus show a weaker evolution of their direction. For an 8.5 Gyrs evolution, we found a difference in the angular separation of the AM vector of around 0.4 in $cos(\alpha)$ regardless of halo mass. 
 
 \item      The main trends reported here for haloes with different numbers of particles appear to be the same. 
            There is a dependence in most of the relations with halo mass, but this dependence is smooth
            with no obvious discontinuities that suggest resolution and/or numerical limitation. 
            Thus, if we apply a more conservative cut in the number of particles than the one used throughout the paper (eg. 40,000 particles as suggested by \citealt{Benson:2017}), 
            the main conclusions of this work would remain unchanged.

\end{itemize}

The results found in this work show that the evolution of the direction of the AM vector is quite complex but that can be well described by the relative mass change and time in which the accretion happens. 
Using that dependence, we constructed look up tables that can be used to produce Monte-Carlo simulations of the evolution of the AM vector of halos in low-resolution simulations. 
This model can be added to any model that makes use of halo merger trees, such as 
SAMs, HODs and SHAMs, regardless of the halo finder algorithm or the minimum number of particles used for identification in their dark matter simulations. This method provides an alternative 
to using direct measurements performed with small numbers of particles ($<1,000$). 
This new method is more accurate as it suffers less from shot noise (to the extent that the mass of halos suffers), 
and is able to describe the evolution of the 
angular momentum amplitude and direction, including AM flips and slews that are important for the evolution of disk sizes and associated phenomena, i.e. star formation rates, disk instabilities, merger driven starbursts and stellar feedback.

\label{Con}

\section*{Acknowledgements}
This work was possible thanks to the efforts of Gerard Lemson and
colleagues at the German Astronomical Virtual Observatory (GAVO) in setting
up the Millennium Simulation database in Garching. 
We thank the people in charge of the CosmoSim web page and the responsible 
of uploading the T-Web to that server.
We thank
Cedric Lacey, Carlton Baugh, Peder Norberg, and Jaime Forero-Romero for many useful discussions. 
We acknowledge support from the European Commission's Framework Programme
7, through the Marie Curie International Research Staff Exchange Scheme 
LACEGAL (PIRSES-GA-2010-269264).
SC further acknowledges support from CONICYT Doctoral Fellowship Programme. 
NP \& SC acknowledge support from a STFC/Newton-CONICYT Fund award (ST/M007995/1 - DPI20140114) and Anillo ACT-1417. 
CL is funded by a Discovery Early Career Researcher Award (DE150100618). This work was supported by a Research Collaboration Award 2016 at the University of Western 
NP is supported by ``Centro de Astronomıa y Tecnologıas Afines'' BASAL PFB-06
and by Fondecyt Regular 1150300. 
Parts of this research were conducted by the Australian Research Council Centre of Excellence for All-sky Astrophysics (CAASTRO), through project number CE110001020.
The calculations for this paper were performed on the 
ICC Cosmology Machine, which is part of the DiRAC-2 
Facility jointly funded by STFC, the Large Facilities 
Capital Fund of BIS, and Durham University and on the Geryon computer at the Center for 
Astro-Engineering UC, part of the BASAL PFB-06, which received additional
funding from QUIMAL 130008 and Fondequip AIC-57 for upgrades.

\bibliography{Biblio}

\appendix

\section{Calculation of $M_*$}
Here we present the procedure to calculate $M_*$, following the work of \citep{RP:2016} for the calculation of $M_C$:

\begin{equation}
\sigma(M_*,a) = 1,
\end{equation}
\noindent
where $a$ is the expansion factor and $\sigma(M)$ is the amplitude of perturbations calculated as
\begin{equation}
 \sigma^2(M) = \left( \frac{D(a)}{D(1)} \right)^2\frac{1}{2\pi^2}\int_0^\infty k^2P(k)W^2(k,M)dk.
\end{equation}
\noindent
Here, $P(k)$ is the power spectrum of perturbations, $W(k, M)$ is the Fourier transform of the real-space top-hat 
filter corresponding to a sphere of mass $M$, and $D(a)$ is the amplitude of the growing mode
\begin{equation}
D(a) \equiv a\ g(a) =\frac{5}{2}\left( \frac{\Omega_{M,0}}{\Omega_{\Lambda,0}} \right)^{1/3}\frac{\sqrt{1+x^3}}{x^{3/2}}\int_0^x\frac{x'^{3/2}}{[1+x'^3]^{3/2}}dx',
\end{equation}

\begin{equation}
x  \equiv \left( \frac{\Omega_{\Lambda,0}}{\Omega_{M,0}} \right)^{1/3}a.
\end{equation}
\noindent
Here, $\Omega_{\Lambda,0}$ and $\Omega_{M,0}$ are the density contributions of matter and the cosmological constant at $z = 0$ respectively, and $g(a)$ is the lineal 
growth factor \citep{Hamilton:2001,Klypin:2011}.

\label{lastpage}
\end{document}